\documentclass[12pt]{article}
\usepackage{amsmath}
\def\R{{\mathcal R}}
\def\be{\begin{equation}}
\def\ee{\end{equation}}

\begin{document}
\title{New Massive Gravity and AdS$_4$ counterterms }
\smallskip\smallskip\smallskip
\author{{ Dileep P. Jatkar}$^{1}$\thanks{email: dileep@hri.res.in} $~~$ and
{ Aninda Sinha}$^{2}$\thanks{email: asinha@cts.iisc.ernet.in}
\\[8mm]
\it $^1$Harish-Chandra Research Institute \\
\it Chhatnag Road, Jhusi, Allahabad, 211019, India\\[2mm]
\it $^2$Centre for High Energy Physics, Indian Institute of Science,\\
\it C. V. Raman Avenue, Bangalore 560012, India}

\maketitle
\vskip2cm \abstract{ We show that the recently proposed DBI extension
  of new massive gravity emerges naturally as a counterterm in
  AdS$_4$. The resulting on-shell Euclidean action is independent of
  the cut-off at zero temperature. We also find that the same choice of counterterm gives the usual area law for the AdS$_4$ Schwarzschild black hole
  entropy in a cut-off independent manner. The parameter values of the
  resulting counterterm action correspond to a $c=0$ theory in the
  context of AdS$_3$/CFT$_2$. We rewrite this theory in terms of the
  gauge field that is used to recast 3D gravity as a Chern-Simons
  theory. }

\newpage
%\section{Introduction}
%\label{sec:introduction}

New massive gravity(NMG) \cite{bht1} is an interesting higher
derivative theory of gravity in 2+1 dimensions and has been the
subject of much interest in recent times.  The propagating degree of
freedom in this theory is a massive spin-2 field. Interestingly, a different derivation based on the AdS/CFT correspondence
was provided in \cite{sinha1}. Assuming that a dual conformal field theory exists, one can demand the existence of a c-theorem which posits the existence of a monotonic function in the space of couplings as the
theory flows from the UV to the IR. It is the existence of a simple 
c-function that made this derivation possible in \cite{sinha1}. 
 By demanding the existence of a simple c-function
\cite{sinha2}, this theory can be extended to more than four
derivative order \cite{sinha1,paulos}.  Of course one can ask if there are any interesting theories which are not truncated at a particular order, since any such truncation
raises the obvious question as to what happens to higher order terms.

An interesting infinite order extension of gravity is Dirac-Born-Infeld (DBI) gravity.  This theory agrees with NMG to
quadratic order\footnote{In fact there were two different actions
  proposed in \cite{tekin1} only one of which agrees with the AdS/CFT
  analysis.} when a weak field expansion is made.  Quite remarkably,
the cubic and quartic terms of this theory agree with the values
obtained from the AdS/CFT calculation leading to a simple c-function
\cite{tekin1,tekinc}.  It was shown in \cite{tekinc}, that the DBI-NMG
theory also admits a simple c-function. Further properties of this
theory have been investigated in \cite{dbinmgrefs}.

Unfortunately, apart from the usual two derivative theory or $f(R)$
theories, it is known that there is a conflict between the unitarity
of the bulk gravity theory and that of the boundary CFT \cite{liu1, bht2,
  tekinu}.  It is quite possible that only the $c=0$ theories are to
be taken seriously in this context.  The relevant CFT may be a
logarithmic CFT.  There has been some evidence to this effect in the
literature \cite{liu1, gruhohm, grumiller1,gaberdiel}.  Although these
higher derivative theories are interesting in their own right, a
fundamental explanation for their origin is still lacking.

In this letter, we will make two interesting observations about
NMG. Firstly, by examining counterterms in AdS$_4$ with different
boundary topologies and demanding that the total Euclidean action is
independent of the cut-off, we will find that the boundary counterterm
coincides with the DBI extension of NMG.  The same  counterterm leads to an area law for entropy for the Schwarzschild black hole in a cut-off independent manner. Further, it  turns out that precisely
for this choice of parameter values the DBI-NMG action in 2+1
dimensions would correspond to a $c=0$ CFT in 1+1
dimensions.  Secondly, we will find a simple way to rewrite this
DBI-NMG action in terms of the gauge field that is used to recast 2+1
dimensional Einstein gravity into a Chern-Simons theory \cite{at,witten}. 

%\section{NMG from counterterms}

It is well known that in order to make sense of the bulk action in the
AdS/CFT correspondence, one needs to add counterterms since the
gravity action typically diverges
\cite{bk,robcounter,krauslarsen,sken}.  In \cite{robcounter} for
example, it was shown that the full (Euclidean) gravitational action
in $D=d+1$ spacetime dimensions has three contributions

\begin{equation}
  \label{eq:actads}
I_{AdS}=I_{bulk}(g_{\alpha\beta})+I_{surf}(g_{\alpha\beta})+I_{ct}(h_{mn})\,,  
\end{equation}
where $I_{bulk}$ is the familiar classical action given by

\begin{equation}
  \label{eq:actblk}
I_{bulk}=-\frac{1}{2\ell_P^{d-1}} \int_{{\mathcal M}} d^{d+1} x
\sqrt{g}\left(R+\frac{d(d-1)}{L^2}\right)\,,  
\end{equation}
$I_{surf}$ is the Gibbons-Hawking term given by

\begin{equation}
  \label{eq:surf}
I_{surf}=-\frac{1}{\ell_P^{d-1}}\int_{\partial {\mathcal M}} d^d x \sqrt{h} K\,,  
\end{equation}
with $K=h^{mn}\nabla_m \hat n_n$ being the trace of the extrinsic
curvature of the boundary.  Here $h_{mn}$ is the induced metric on
the boundary defined through $h_{mn}=g_{mn}-\hat n_m \hat n_n$ with
$\hat n$ being an outward pointing unit normal vector to the boundary. The
counterterm action $I_{ct}$ can be arranged as an expansion in powers
of the boundary curvature \cite{robcounter}:
\begin{eqnarray}\label{robeq}
I_{ct}&=&\frac{1}{\ell_P^{d-1}}\int_{\partial {\mathcal M}} d^d
x\sqrt{h}\left[\frac{d-1}{L}+\frac{L}{2(d-2)} \R \right. \nonumber\\
&& ~~~~~~~~~\left.+\frac{L^3}{2(d-4)(d-2)^2}\left(\R_{\mu\nu}\R^{\mu\nu}-
    \frac{d}{4(d-1)}\R^2\right)+\cdots\right]\,, 
\end{eqnarray}
where $\R$ and $\R_{\mu\nu}$ are the Ricci scalar and Ricci tensor made
out of $h_{\mu\nu}$. Our interest in this paper is $d=3$ in other words
the counterterms in AdS$_4$. Notice that the four derivative terms in
eq.(\ref{robeq}) take the form
\begin{equation}
  \label{eq:rsq}
\R_{\mu\nu}\R^{\mu\nu}-\frac{3}{8}\R^2\,,  
\end{equation}
which is {\it precisely} what arises in new massive gravity! Let us
review how this comes about in the analysis of \cite{robcounter}. AdS
metrics can be written as

\begin{equation}\label{adsm}
ds^2=\frac{dr^2}{k+\frac{r^2}{L^2}}+(k+\frac{r^2}{L^2})
d\hat\Sigma^2_{-k,\hat m}+\frac{r^2}{L^2}d\tilde \Sigma^2_{k,\tilde
  m}\,, 
\end{equation}
where $k=0,\pm 1$. The metrics $d\hat\Sigma^2_{-k,\hat m}$ and
$d\tilde \Sigma^2_{k,\tilde m}$ are defined through
\begin{equation}
  \label{eq:met}
d\Sigma^2_{k,m}=\begin{cases} L^2 d\Omega_m^2  & \mbox{for~} k=+1\\
                              \sum_{i=1}^m dx_i^2 & \mbox{for~} k=0 \\
                              L^2 d\Psi_m^2  & \mbox{for~} k=-1
                \end{cases}
\end{equation}
where $d\Omega_m^2$ is the metric on a unit $m$-sphere and $d\Psi_m^2$
is the metric on a $m$-hyperboloid whose metric is obtained by the
analytic continuation of the metric on the unit $m$-sphere. In this
way AdS$_4$ can be written such that the boundary is on $R^3, S^3,
H^3, S^2 \times S^1, H^2 \times S^1$. Consider the case when the
boundary is $S^3$. To derive the above form for $I_{ct}$ let us begin
with
\begin{equation}
  \label{eq:ct}
I_{ct}=\frac{1}{\ell_P^2} \int_{\partial {\mathcal M}} d^3 x
\sqrt{h}\left(\frac{\kappa}{L}+\mu L \R+L^3 (\lambda_1
  \R_{\mu\nu}\R^{\mu\nu}+\lambda_2 \R^2)\right)\,.   
\end{equation}
Then integrating $r$ between $0$ and $\Lambda$, with $\Lambda$ being a
cut-off, we find 
\begin{equation}
  \label{eq:blk}
  I_{bulk}+I_{surf}=2 \frac{\Lambda^2
  V_{S^3}}{\ell_P^2}(1+\frac{\Lambda^2}{L^2})^{3/2}\,. 
\end{equation}
The square-root form of the result will give us motivation to consider
a compact {\it all-order} expression for the counterterm to be made precise shortly. For
the moment, we find that expanding around $\Lambda\rightarrow \infty$
gives us
\begin{equation}
  \label{eq:actads3}
  \frac{I_{AdS, S^3}}{V_{S^3}}=\frac{\Lambda^3}{L
  \ell_P^2}(2-\kappa)+\frac{\Lambda L}{\ell_P^2} (3-6\mu)+\frac{3
  L^3}{4 \ell_P^2 \Lambda}(1-16\lambda_1-48\lambda_2)+\cdots\,. 
\end{equation}
We can do a similar analysis for $S^1\times S^2$ (or $S^1 \times H^2$)
which gives us 
\begin{equation}
  \label{eq:adss2}
  \frac{I_{AdS, S^1\times S^2}}{V_{S^1\times S^2}}=\frac{\Lambda^3}{L
  \ell_P^2}(2-\kappa)+\frac{\Lambda L}{2\ell_P^2}
(\kappa-4+4\mu)+\frac{L^3}{8 \ell_P^2
  \Lambda}(\kappa-16\lambda_1-32\lambda_2-8\mu)+\cdots\,. 
\end{equation}
Then demanding that the $\Lambda^3$ and $\Lambda$ divergences cancel
gives us $\kappa=2, \mu=1/2$. Now let us demand that the final answer
is independent of the cut-off $\Lambda$. This means that the
$1/\Lambda$ terms should also cancel. This leads to $\lambda_1=-1/2,
\lambda_2=3/16$ which is what eq.(\ref{robeq}) has. This form for
$I_{ct}$ makes the result cut-off independent upto $O(1/\Lambda)$ for
any form for AdS allowed by eq.(\ref{adsm}).

The square-root form for $I_{bulk}+I_{surf}$ for the $S^3$ case
makes it very tempting to conjecture a square-root form for the
counterterm action. In particular, let us consider
\begin{equation} \label{eqm1}
I_{ct}=\beta_1 \sqrt{-{\rm det}(\R_{\mu\nu}+\alpha_1 \R
  h_{\mu\nu}+\frac{\alpha_2}{L^2} h_{\mu\nu})}\,.
\end{equation}
Mann in \cite{mann} considered a similar counterterm without the
$\R_{\mu\nu}$ to remove divergences in AdS$_4$.
We find that if the total action $I_{AdS}$ is cut-off independent
to all orders in $\Lambda$ for $S^3$ then there are two choices:

Either
\begin{equation} \label{mainp}
\alpha_1=-\frac{1}{2}\,,\quad \beta_1=-\frac{2L^2}{\ell_P^2}\,,\quad
\alpha_2=-1\,, 
\end{equation}

or
\begin{equation}
\alpha_1=-\frac{1}{6}\,,\quad \beta_1=\frac{2i L^2}{\ell_P^2}\,,\quad
\alpha_2=1\,. 
\end{equation}
While the first choice gives us precisely the DBI gravity action
proposed in \cite{tekin1}, the second one\footnote{The second choice
  is presumably relevant for de Sitter spaces.}  is also an action
proposed in \cite{tekin1} to give NMG upto quadratic order. The
actions in \cite{tekin1} have an extra free coupling which we will
address in a moment. This seems like a remarkable coincidence!  A
comment we would like to make here is that we do not have the freedom
to add terms like $\R_{\rho\sigma}\R^{\rho\sigma} h_{\mu\nu}$, $\R^2
h_{\mu\nu}$ or $\R_{\mu\rho}\R^{\rho\sigma}h_{\sigma\nu}$ to
eq.(\ref{eqm1}) since these terms behave like $1/r^2$ and would spoil
the exact cancellation of the cut-off dependence.

Performing the same analysis for $S^1 \times S^2$ selects
eq.(\ref{mainp}) as the only set of parameters. As shown in
\cite{tekin1, tekinc} this choice is also consistent with the
$c$-theorem method of extending the NMG actions to higher orders in
\cite{sinha1}.  The counterterm eq.(\ref{eqm1}) with eq.(\ref{mainp})
give us a cut-off independent result for any metric of the form in
eq.(\ref{adsm}). 

If we look at the counterterm action
in eq.(\ref{eqm1}), it is tempting to attempt a field redefinition by 
which we can set 
\begin{equation}
  \label{eq:21}
  \tilde h_{\mu\nu} = \R_{\mu\nu}- \frac{1}{2} \R
  h_{\mu\nu}-\frac{1}{L^2} h_{\mu\nu}\, .
\end{equation}
This would reset the entire counterterm action to the non-dynamical
boundary cosmological constant term.  Since $h_{\mu\nu}$ is the
boundary value of the bulk metric, this boundary field redefinition
would need to arise from a more complicated bulk field
redefinition\footnote{It is not clear if a local bulk field
  redefinition leading to eq.(\ref{eq:21}) exists.} which would make
the bulk action higher derivative. Hence we will refrain from making
this redefinition. 

At this point, one can ask if these counterterms give rise to sensible
thermodynamics for AdS black holes even in the presence of a cut-off.
It would be a strong check if we can show that the Bekenstein-Hawking
area law is still satisfied in a cut-off independent manner with this
choice of counterterm. This must be true as the bulk action is just Einstein-Hilbert with a cosmological constant. A straightforward, if somewhat tedious,
calculation shows that the on-shell action is cut-off dependent. 
This is actually expected since in the free energy expression, entropy times temperature appears. If the temperature is cut-off dependent, as turns out to be the case,
it must be that to have a cut-off independent entropy, the free energy and hence the action must be cut-off dependent.
% This may at first sight seem worrisome, however, the
%on-shell action is not the physical quantity of relevance here.  One
%should rather calculate the entropy from this action.  
At this stage,
it is essential to realize that the definition of temperature needs
some care.  The point is that we need to use a coordinate system that
ensures that the speed of light of the boundary theory (with cut-off)
is unity.  This entails redefining the time coordinate by a cut-off
dependent factor which in turn renormalizes the temperature.  Once this
is taken into account, we indeed find that the entropy is given by the
Bekenstein-Hawking area law as it should!  Let us illustrate this in
more detail for the spherical case.  We write the metric as
\begin{equation}
ds^2=\frac{dr^2}{1+f(r)}+g(r)(d\theta^2+\sin^2\theta d\phi^2)+(1+f(r))
N dt^2\,,
\end{equation}
where we have put in a factor of $N$ in front of $dt^2$.  This factor
is chosen to be $g(\Lambda)/[1+f(\Lambda)]$ to ensure that the speed of
light is unity in the cut-off field theory.  Here $f(r)=r^2/L^2-\mu/r$
and $g(r)=r^2$.  We can write $\mu=r_0+r_0^3/L^2$ with $r_0$ being the
location of the horizon.  Then the temperature works out to be
\begin{equation}
T=\frac{3 r_0^2+L^2}{4 \pi r_0 L^2}\sqrt{N}\,.
\end{equation}
Then we proceed as usual by defining the free energy through $W= T
I_E$ and entropy through $S=-\partial W/\partial T=-\partial_{r_0}
W/\partial_{r_0} T$ where we are holding the cut-off fixed.  This
readily gives
\begin{equation}
S=\frac{8\pi^2 r_0^2}{\ell_P^2}=\frac{A}{4G}\,.
\end{equation}
Without the crucial $N$ dependent factor in the definition of
temperature, one would not get this expected result unless we are in
the $\Lambda=\infty$ limit.  This computation therefore demonstrates
that our counterterm is indeed doing its role as expected in the
computation of the Bekenstein-Hawking entropy.  Writing the entropy density as $s= c_S T^2$
then would imply that the effective number of degrees of freedom, $c_S$, depends on the
cut-off.
%% {\bf AS: The last couple of statements should be taken with a
%%   barrel of salt at this stage} 

Let us now ask the question as to what happens if $I_{BI}$ was treated as a stand-alone action rather than a counterterm. 
In the context of AdS/CFT, $I_{BI}$ will be dual to a 1+1 CFT. 
The action proposed in \cite{tekin1} in Euclidean signature consists
of two pieces
\begin{equation}
  \label{eq:bi}
  I_{BI}=-\frac{4m^2}{\kappa^2}\int d^3 x\left[\sqrt{-\det
    \left(\frac{{\mathcal G}}{m^2}\right)}-(\frac{\lambda}{2
    m^2}+1)\sqrt{\det h}\right]\,,
\end{equation}
where ${\mathcal G}_{\mu\nu}=\R_{\mu\nu}-\frac{1}{2} h_{\mu\nu} \R-m^2
h_{\mu\nu}$.  In order for the weak field expansion to be consistent
with eq.(\ref{robeq}), we need to set $\lambda=2/L^2$. But then
comparing with eq.(\ref{eqm1}) and eq.(\ref{mainp}) we have
$(\frac{\lambda}{2 m^2}+1)=0$.  
 As
shown in \cite{tekinc}, the central charge of the 1+1 CFT is
proportional to $(\frac{\lambda}{2 m^2}+1)$. Thus the choice of
parameters in eq.(\ref{mainp}) in the AdS/CFT context corresponds to
having a dual CFT with zero central charge! Of course this only makes
sense if we treated this action as a stand-alone action in the context
of AdS$_3$/CFT$_2$ and not as a counterterm to AdS$_4$. Further, the
mass of the massive mode has been pushed to zero {\footnote{In a previous version, it was erroneously stated that the mass is pushed to infinity. We thank B. Tekin for pointing this out.}}.  It will be
interesting to consider fluctuations of the boundary metric along the
lines of \cite{compere} with Neumann boundary conditions.

One important comment about the stand-alone $c=0$ DBI-NMG action is the following. Consider doing the 
field redefinition in eq.(\ref{eq:21}). Naively this would appear to
make the theory trivial.  
However, the field redefinition cannot trivialise this
action. In order to see this,  consider the path
integral representation for this theory.  While the field
redefinition in eq.(\ref{eq:21}) would reduce the
action to a trivial volume element, the change in the integration
measure will give rise to a Jacobian factor.  Since the field
redefinition involves derivatives of the metric tensor, the Jacobian
when put back in the exponent will reinstate the dynamics into the new
action\footnote{We thank Rob Myers for pointing this out to us.}.

It is well known that three dimensional gravity can be written in
terms of a Chern-Simons gauge theory \cite{at,witten}.  The one form
gauge field is a linear combination of spin connection $\omega_{ab}$
and dreibein,
\begin{equation}
  \label{eq:1}
  A^{a\pm} = \omega^a \pm \frac{1}{\ell} e^a\, ,
\end{equation}
where $\omega^a = \epsilon^{abc}\omega_{bc}/2$ is dualised spin
connection and $e$ is dreibein and $a$ is a gauge index corresponding
to SL(2,R)\footnote{We follow the conventions in \cite{carlip} and
  these are summarized in the appendix.}.  Note that $a$ runs from 1
to 3 which coincides with the number of spacetime dimensions.  The
field strength corresponding to this gauge field can then be related
to the Riemann curvature subject to metric compatibility of the
connection,
\begin{equation}
  \label{eq:2}
%%   F^{a\pm}_{\mu\nu} = \frac{1}{2}\epsilon^{abc} R_{\mu\nu}^{~~\rho\sigma}
%%   e^b_\rho e^c_\sigma + \frac{1}{2 \ell^2} \epsilon^{abc} e^b_\mu
%%   e^c_\nu \pm \frac{1}{\ell}  \left(\partial_\mu e^a_\nu  - \partial_\nu e^a_\mu
%%  + \epsilon^{abc}\omega^b_\mu e^c_\nu\right)\, .
F^{a\pm} = \frac{1}{2}\epsilon^{abc} R^{bc} + \frac{1}{2 \ell^2}
\epsilon^{abc} e^b \wedge e^c \pm \frac{1}{\ell}  \left( d e^a +
  \epsilon^{abc} \omega^b\wedge e^c \right)
\end{equation}
The metric compatibility condition in terms of dreibein and spin
connection is the torsion free constraint
\begin{equation}
  \label{eq:6}
%%   \partial_\mu e^a_\nu  - \partial_\nu e^a_\mu +
%%   \epsilon^{abc}\omega^b_\mu e^c_\nu = 0\, .
d e^a + \epsilon^{abc} \omega^b\wedge e^c = 0\, .
\end{equation}
Once we impose the torsion free constraint, the field strengths for both
gauge fields is identical because before imposing the constraint they
differ by the torsion constraint.  For later purposes it is
appropriate to define the field strength 
\begin{equation}
  \label{eq:16}
  \mathcal{F}^a = \frac{1}{2} (F^{a+} +
  F^{a-})\, .
\end{equation}
$\mathcal{F}^a_{\mu\nu}$ does not contain the torsion constraint and
therefore is suitable for expressing the curvature tensor in terms of
gauge field strength. 
We will be using eq.(\ref{eq:16}) to rewrite the DBI-NMG action in
terms of gauge fields.  To do that we note the following identity in 3
dimensions \cite{carlip} 
\begin{equation}
  \label{eq:4}
  G^\mu_\nu = -
  \frac{1}{4}\varepsilon^{\mu\rho\sigma}\varepsilon_{\nu\lambda\kappa}
  R_{\rho\sigma}^{~~\lambda\kappa} \, .
\end{equation}
Since we are interested in writing all geometric quantities in terms
of gauge fields, we will use eq.(\ref{eq:2}) and eq.(\ref{eq:16}) to
express the Einstein tensor in terms of the gauge field strength
$\mathcal{F}^a_{\mu\nu}$ 
\begin{equation}
  \label{eq:5}
  G^\mu_\nu = - \frac{1}{4}
  \varepsilon^{\mu\rho\sigma}\varepsilon_{\nu\lambda\kappa} ( \epsilon^{abc}
  \mathcal{F}^a_{\rho\sigma} e^{b\lambda}e^{c\kappa} - \frac{1}{\ell^2}
  \delta^\lambda_\rho \delta^\kappa_\sigma)\, .
\end{equation}
It is more instructive to write the  Einstein tensor with mixed indices
\begin{equation}
  \label{eq:7}
  G^{a\mu} = - \frac{1}{2\sqrt{\det h}}\varepsilon^{\mu\nu\rho} (
  \mathcal{F}^a_{\nu\rho} - 
  \frac{1}{\ell^2} \epsilon^{abc} e_{b\nu}e_{c\rho})\, .
\end{equation}
This equation can be further rearranged to write the gauge field
strength in terms of geometric quantities, namely the dreibeins and
the Einstein tensor,
\begin{equation}
  \label{eq:8}
  G^{a\mu}- \frac{1}{2\ell^2\sqrt{\det h}}
  \varepsilon^{\mu\nu\rho}\epsilon^{abc} e_{b\nu}e_{c\rho} = - 
  \frac{1}{2\sqrt{\det h}}\varepsilon^{\mu\nu\rho}
  \mathcal{F}^a_{\nu\rho} \equiv \frac{1}{\sqrt{\det h}}
  \star\mathcal{F}^{a\mu}\, , 
\end{equation}
using which we find
\begin{equation}
  \label{eq:11}
  \sqrt{-\det(\mathcal{G})} = \sqrt{\det{\star \mathcal{F}^{a\mu}}}\, . 
\end{equation}
The dual gauge field strength carries mixed indices,
however as mentioned before, in three dimensions, gauge and spacetime
indices have same 
range and therefore the determinant of dual field strength is a sensible
quantity to consider.  In fact, it can be written in a more suggestive form as
\begin{equation}
  \label{eq:12}
  \sqrt{-\det(\mathcal{G})} = 
  \sqrt{\det{\star \mathcal{F}}} = 
  \sqrt{\frac{1}{6}\varepsilon^{\mu\nu\rho}\epsilon^{abc}
    \star\!\mathcal{F}_{a\mu} \star\!\mathcal{F}_{b\nu}
    \star\!\mathcal{F}_{c\rho}}\, . 
\end{equation}
While the lagrangian density, in terms of the metric, reproduces the
lagrangian density of the new massive gravity in the weak
field expansion, there is no such expansion available in the gauge
field formulation.

At this point, it is interesting to note that we can write a more
general action by adding a Chern-Simons action to this gauge theory
action:
\begin{eqnarray}
  \label{eq:13}
  I_{gen} &=& \xi \int d^3x\, {\rm Tr}\left[
  A^{+}\wedge dA^{+} + A^{+}\wedge A^{+}\wedge A^{+} \right.\nonumber \\
 &-& \left. A^{-}\wedge dA^{-} - A^{-}\wedge A^{-}\wedge A^{-}
  \right] + \eta\int d^3x\, \sqrt{\det \star\mathcal{F}}\, .
\end{eqnarray}
%% where $k = \ell/8G_N$ and $\kappa^2 = 8\pi G_N$.  
We can express (\ref{eq:13}) in terms of the gravity variables,
\begin{equation}
  \label{eq:14}
  I_{gen} = \frac{\xi}{\ell}\int d^3x \sqrt{\det h}\left[ R +
    \frac{1}{\ell^2}\right] + 
    \eta \int d^3x \sqrt{-\det\mathcal{G}} \, .
\end{equation}
This would correspond to a 1+1 CFT with $c\neq 0$ for non-vanishing $\xi$.

In this letter, we have found that the DBI extension of new massive gravity can arise as a counterterm in AdS$_4$. This choice makes the zero temperature action cut-off independent and gives a sensible area law for the Schwarzschild black hole. 
Let us now comment on possible generalizations to higher
dimensions.  If one considers odd bulk dimensions, then the dual CFT is
even dimensional and has conformal anomalies.  These manifest
themselves as logarithmic terms in the on-shell bulk action evaluated
for non-trivial boundary topologies.  There are no local counterterms
which can remove these logarithmic divergences and it is not apriori
clear that the magic in 3+1 dimensions can be extended to odd
dimensional bulk theories.  We leave the examination of higher even
bulk dimensions as an interesting open problem. It will also be interesting to work out the counterterms in the presence of higher derivative terms in the bulk action (see for example \cite{hd} for interesting actions with higher derivative terms) to see if there is a general lesson to be learnt.

Finally let us comment on the relation between recent work on
holographic renormalization group flows \cite{strom,son,pol,ranga}.
These attempt at putting the boundary at a finite radius and
integrating out all the degrees of freedom associated with larger
values of radius.  Schematically the total action is written as
\cite{ranga} \be S=\int_{r<\Lambda} d^{d+1}x \sqrt{g} {\mathcal L}+
S_B \ee where $S_B$ is a boundary action defined on $r=\Lambda$ and
can be viewed as a boundary state for the bulk theory in the region
$r<\Lambda$.  The choice of the cut-off is arbitrary and the on-shell
action is independent of the cut-off in case of pure AdS bulk.
However, in case of geometries which asymptote to AdS, the on-shell
action is cut-off dependent but it is just as well, because the
cut-off dependence of the on-shell action as well as the local
temperature is such that it correctly reproduces the
Bekensetin-Hawking entropy in case of AdS-black hole backgrounds.  In
what we are doing $S_B$ is the sum of the Gibbons-Hawking term and the
counterterm.  Since we have found the total action to be cut-off
independent for $T=0$, we have effectively solved the flow equations
in \cite{ranga} for gravity to leading order in the context of
AdS$_4$/CFT$_3$.  
%% In order to solve the full problem, one would need
%% to find a set of counterterms that would leave the total action
%% cut-off independent for an arbitrary background which seems to be a
%% much harder problem.  
%% For instance, with the DBI counterterm, the
%% stress tensor which is effectively the one point function is not
%% cut-off independent.  We will leave a more thorough analysis of this
%% problem for future work.  

To understand utility of these counterterms, it is instructive to look
at them from the boundary field theory point of view\footnote{We thank
  Ashoke Sen for discussions on this issue.}.  In the field theory, one
defines a set of local counterterms once and for all, irrespective of
what vacuum one is working with.  One could then be working with a
cut-off field theory or with a finite temperature field theory--this
does not affect the local counterterms.  For example, the Euclidean
description of a finite temperature field theory corresponds to making the
Euclideanised time coordinate periodic.  This global periodicity
condition does not affect the form of the local counterterms.  As a result
even thermal field theories will have the same set of local counterterms.
Of course, one has to ask appropriate questions about physical
observables in such theories.  We have looked at this problem from the
bulk side and have demonstrated that the Bekenstein-Hawking entropy of
the black hole in AdS space can be recovered if we define the local
temperature at the cut-off surface appropriately. In the same vein,  it would be
interesting to find a prescription for correlation functions of
physical observables.  We will leave this analysis for future
work. 

\vskip 5mm

{\bf Acknowledgments} : We would like to thank P. Argyres,
G. Arutyunov, N. Banerjee, S. R. Das, B. de Wit, S. Dutta,
R. K. Gupta, J. Hung, M. Paulos,  M. Rangamani, A. Shapere, B. Tekin,
S. Theisen and S. Vandoren for discussions.  In particular, we would
like to thank R. Myers and A. Sen for useful suggestions and
discussions and R. Myers, M. Rangamani and B. Tekin for comments on the
draft.  We would like to thank Perimeter Institute for hospitality
during this work.  D.P.J would like to thank University of Kentucky,
Utrecht University, Albert Einstein Institute and Indian Institute of
Science for hospitality during the course of this work.  A.S. thanks
Princeton University for hospitality where part of this work was
presented.  We thank the organizers for ISM2011 for hospitality and an
opportunity to present this work.
  
\section*{Appendix}
\label{sec:appendix}

Here we will summarise our conventions.  We will be working in the
Euclidean signature.  We use indices  $\mu, \nu,
\rho, \cdots$ to denote three dimensional and $a, b, c, \cdots$ to
denote gauge/local Lorentz indices in three dimensions.  Indices
$\alpha, \beta, \cdots$ are general $d$-dimensional indices.

The $SL(2,R)$ generators are given by
\begin{equation}
 \label{eq:17}
 T^0 = \frac{1}{2}\left(
   \begin{array}{cc}
     0 & -1 \\ 1 & 0
   \end{array}
\right), \,\,
T^1 = \frac{1}{2}\left(
   \begin{array}{cc}
     1 & 0 \\ 0 & -1
   \end{array}
\right), \,\,
T^2 = \frac{1}{2}\left(
   \begin{array}{cc}
     0 & 1 \\ 1 & 0
   \end{array}
\right)\, ,
\end{equation}
such that
\begin{equation}
 \label{eq:18}
 {\rm Tr}(T^a T^b) = \frac{1}{2} \eta^{ab},\quad [T^a, T^b] =
 \epsilon^{abc} T^c\, ,
\end{equation}
where, $\epsilon^{012} = 1$.  The group indices $a,b,c,\cdots$ are
raised and lowered by the flat Lorentzian metric $\eta_{ab}$.  The
Levi-Civita tensor $\varepsilon_{\mu\nu\rho}$ is totally antisymmetric
tensor with $\pm1$ and $0$ entries.  The Levi-Civita tensor density
will be written in terms of $\varepsilon_{\mu\nu\rho}$ with explicit
insertions of $\sqrt{\det h}.$

\end{document}